# Multi-star Turbulence Monitor: A new technique to measure optical turbulence profiles

Paul Hickson,[1,2]⋆ Bin Ma,[2,3] Zhaohui Shang,[2] Suijian Xue[2]
[1] *University of British Columbia, Department of Physics and Astronomy, 6224 Agricultural Road, Vancouver, B.C., V6T 1Z1, Canada*
[2] *National Astronomical Observatories, Chinese Academy of Sciences, 20A Datun Road, Chaoyang District, Beijing 100101, China*
[3] *School of Astronomy and Space Science, University of Chinese Academy of Sciences, Beijing 100049, China*

**ABSTRACT**

The strength and vertical distribution of atmospheric turbulence is a key factor determining the performance of optical and infrared telescopes, with and without adaptive optics. Yet, this remains challenging to measure. We describe a new technique using a sequence of short-exposure images of a star field, obtained with a small telescope. Differential motion between all pairs of star images is used to compute the structure functions of longitudinal and transverse wavefront tilt for a range of angular separations. These are compared with theoretical predictions of simple turbulence models by means of a Markov-Chain Monte-Carlo optimization. The method is able to estimate the turbulence profile in the lower atmosphere, the total and free-atmosphere seeing, and the outer scale. We present results of Monte-Carlo simulations used to verify the technique, and show some examples using data from the second AST3 telescope at Dome A in Antarctica.

## 1 INTRODUCTION

Atmospheric turbulence causes random fluctuations in the atmospheric index of refraction, which give rise to the familiar blurring and image motion known as "seeing". This not only impacts seeing-limited observations, but also observations using adaptive optics (AO, Beckers 1993a) In fact, the impact of seeing on telescope performance can be even greater when AO is employed (Racine 2009). Much effort has been devoted to the measurement of atmospheric parameters at astronomical sites, and a variety of techniques have been developed for this purpose (Hickson 2014 provides a short review). A cornerstone method for the measurement of seeing is the differential-image-motion monitor (DIMM, Sarazin & Roddier 1990) which measures the integral of the index of refraction structure constant, $C_n^2$, along the line of sight. However, the vertical distribution of $C_n^2$, is also important, particularly for AO applications. The distribution of turbulence in the upper atmosphere is relevant for the selection of the conjugate height of deformable mirrors in multi-conjugate adaptive optics system (MCAO, Beckers 1988), while the turbulence distribution in the lower km of the atmosphere is important for ground-layer adaptive optics (GLAO, Rigaut 2002; Tokovinin 2004). Turbulence profile measurements can also help establish the optimal location and height above ground for new telescopes.

A variety of techniques have been developed for nighttime measurements of turbulence profiles. These include tower or balloon-borne microthermal measurements (Trinquet et al. 2008), Scintillation Detection and Ranging (SCIDAR, Vernin & Roddier 1973; Masciadri et al. 2010), Slope Detection and Ranging (SLODAR Wilson 2002), and lunar limb observations (Maire et al. 2007). Also, the Multiaperture Scintillation Sensor (MASS) measures turbulence in six high-altitude layers (Tokovinin 1998; Kornilov et al. 2003). When combined with DIMM, it can measure the integrated turbulence in the ground layer (GL).

Techniques that give good resolution within the GL are Sonic Detection and Ranging (SODAR Little 1969; Travouillon 2006; Bonner et al. 2010), Shadow-Band Ranger (SHABAR Beckers 1993b; Beckers et al. 1997; Hickson & Lanzetta 2004; Hill et al. 2006; Tokovinin et al. 2010; Lombardi et al. 2010; Hickson et al. 2013), Generalized SCIDAR (G-SCIDAR, Fuchs et al. 1998; Klueckers et al. 1998), and Surface-layer SLODAR (Osborn et al. 2010).

These techniques all have advantages and limitations, and are in many ways complimentary. Site-test campaigns often employ multiple methods and observations (e.g.

⋆ E-mail: hickson@phas.ubc.ca

Schöck et al. 2009). In some cases, site characteristics or logistical considerations favour particular techniques. The Multi-star Turbulence Monitor (MTM) technique, described in this paper, was motivated by the availability at Dome A in Antarctica of the Antarctic Survey Telescopes (AST3), three robotic wide-field 0.5-m telescopes equipped with high-speed wide-field imagers (Cui et al. 2008; Yuan & Su 2012; Shang et al. 2016; Liu et al. 2018; Ma et al. 2018). However, the technique could have broader application. All that is required is a small telescope, having a well-corrected field of view of half a degree or more, and a digital imager that has an electronic shutter capable of exposure times in the 10 - 100 millisecond range. These are readily available commercially.

In this paper, we develop the theoretical basis for the MTM and describe the data analysis steps. The technique is verified by means of Monte-Carlo simulations, and is illustrated with examples using data obtained at Dome A with the second AST3 telescope. A full analysis of all available data from this telescope, and the implications for astronomical observations at Dome A, will be the subject of a future publication (Ma et al., in preparation).

## 2  THEORY

Consider a series of short-exposure images (frames) of a star field, formed by a telescope that has a circular aperture of diameter $D$. As a result of atmospheric turbulence, the positions of the individual stars will vary slightly from frame to frame. The variance of the differential angular displacement of the centroid positions of two stars separated by an angle $\alpha$ is a continuous function of $\alpha$, known as the structure function $\mathcal{D}(\alpha)$ of angular displacement. The displacement can be resolved into longitudinal and transverse components representing image motion that is parallel and perpendicular, respectively, to the vector connecting the two stars in the image plane. We denote these components by the subscripts $+$ and $-$ respectively. As these directions are orthogonal, the respective fluctuations are independent, so the structure function for total displacement can be resolved into the sum of structure functions for the individual components,

$$\mathcal{D}(\alpha) = \mathcal{D}_+(\alpha) + \mathcal{D}_-(\alpha). \tag{1}$$

These structure functions can be found by integrating the turbulence power spectrum multiplied by the appropriate filter functions. For the difference between two beams at angle $\alpha$, the variance of differential phase is given by Eqn 2.123 in Sasiela (2007). (It is necessary to assume that the turbulence is sufficiently weak that the first-order Rytov approximation for the wavefront phase distortion is valid, which is always the case for astronomical observations.) For our application this becomes

$$\mathcal{D}_\pm = 0.2073\, k^2 \int_0^\infty dz\, C_n^2(z) \int d^2\kappa f(\kappa) \cos^2(\kappa^2 z/2k)$$
$$\times F_\pm(\boldsymbol{\kappa}) 2[1 - \cos(\kappa_+ z\alpha)]. \tag{2}$$

Here $z$ is the line-of-sight distance from the observer to the turbulence, $k$ is the optical wavenumber, related to the wavelength $\lambda$ by $k = 2\pi/\lambda$, $\boldsymbol{\kappa}$ is the two-dimensional spatial frequency, having longitudinal component $\kappa_+$. The function $f(\kappa)$ expresses the $\kappa$ dependence of the power spectrum of atmospheric index-of-refraction fluctuations, assumed to be isotropic. We have assumed here that $\alpha \ll 1$, so that the separation between the two beams, at distance $z$, can be represented by $\alpha z$.

As we are interested in the centroid positions of the star images, without AO correction, we use the filter functions for G-tilt. These are given by Eqn 3.38 in Sasiela (2007),

$$F_+(\boldsymbol{\kappa}) = \left(\frac{4}{kD}\right)^2 J_1^2(\kappa D/2) \cos^2\phi, \tag{3}$$

$$F_-(\boldsymbol{\kappa}) = \left(\frac{4}{kD}\right)^2 J_1^2(\kappa D/2) \sin^2\phi, \tag{4}$$

where $\phi$ is the polar angle in the $\boldsymbol{\kappa}$ plane, measured from the longitudinal direction, $D$ is the aperture diameter and $J_1$ is a Bessel function of the first kind.

In Eqn. (2), the $J_1^2$ factors become small when $\kappa \gtrsim 2/D$. Therefore, if $z \ll D^2 k/2 \sim D^2/\lambda$, the diffraction term, $\cos^2(\kappa^2 z/2k)$, will be close to unity. This is the case for apertures greater than 0.1 m at visible wavelengths. Substituting Eqns. (3) and (4) into Eqn. (2), setting the diffraction term equal to unity, and integrating over $\phi$, we obtain

$$\mathcal{D}_\pm(\alpha) = 20.84\, D^{-2} \int_0^\infty dz\, C_n^2(z) \int_0^\infty d\kappa \kappa f(\kappa) J_1^2\left(\frac{\kappa D}{2}\right)$$
$$\times [1 - J_0(\kappa z\alpha) \pm J_2(\kappa z\alpha)], \tag{5}$$

### 2.1  Kolmogorov spectrum

It is instructive to first consider the Kolmogorov spectrum, $f(\kappa) = \kappa^{-11/3}$. The $\kappa$ integrals can be evaluated in terms of generalized hypergeometric functions (Appendix A.1). The results are

$$\mathcal{D}_\pm = \frac{20.84}{D^{1/3}} \int_0^\infty dz\, C_n^2 I(x) \tag{6}$$

where

$$I(x) = \begin{cases} I_1 - S_{10} \pm S_{12}, & x > 1 \\ I_1 - S_{20} - S_{30} \pm (S_{22} + S_{32}), & x < 1 \end{cases} \tag{7}$$

The functions $I_1, S_{10}, S_{12}, S_{20}, S_{22}, S_{30}$ and $S_{32}$, defined in Appendix A.1, depend only on the dimensionless parameter $x = \alpha z/D$.

It is simplest to consider a single thin turbulent layer. The response for extended turbulence can be regarded as a sum over thin layers, weighted by $C_n^2(z)$. Fig. 1 shows the mean square differential image motion as a function of angular separation, for a thin layer at distance $z$. For small angles the variances increase quadratically with $\alpha$. Thus the RMS differential image motion increases linearly with angular separation. However when $\alpha \gtrsim D/z$, differential image motion increases less rapidly.

### 2.2  von Kármán spectrum

For the von Kármán turbulence spectrum,

$$f(\kappa) = (\kappa^2 + \kappa_0^2)^{-11/6}, \tag{8}$$

where $\kappa_0 = 2\pi/L_0$ and $L_0$ is the outer scale of turbulence. The integrals needed to compute the structure functions for this case are evaluated in Appendix A.2. The result can be written in the form

$$\mathcal{D}_\pm = \frac{20.84}{D^{1/3}} \int_0^\infty dz\, C_n^2 [I_3 - I_{40} \pm I_{42}], \tag{9}$$

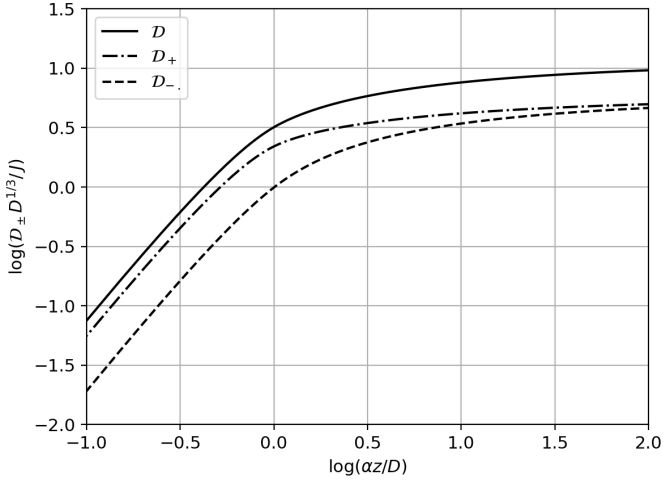

**Figure 1.** Mean square differential image motion for a thin turbulent layer at distance $z$, for a Kolmogorov turbulence spectrum.

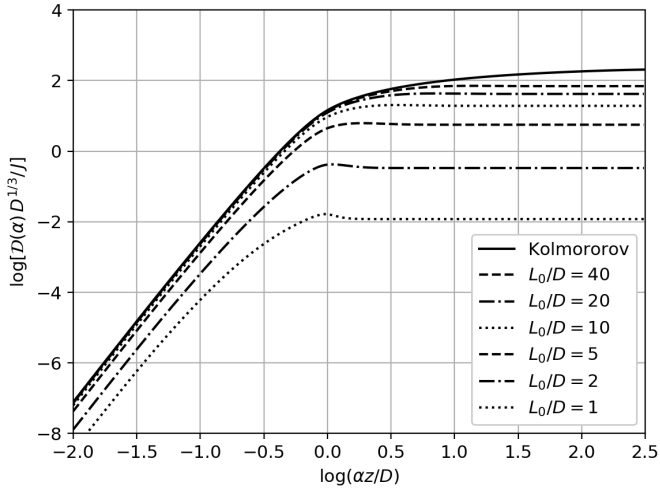

**Figure 2.** Mean square differential image motion for a thin turbulent layer at distance $z$, for a von Kármán turbulence spectrum.

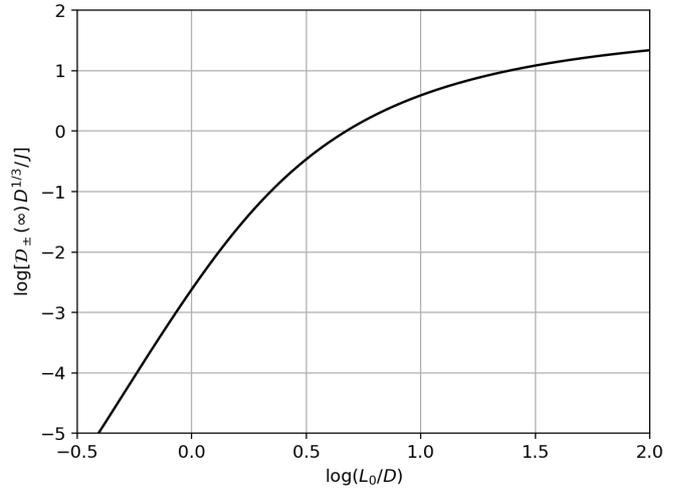

**Figure 3.** Limiting mean square differential image motion as a function of the outer scale.

where $I_3$, $I_{40}$ and $I_{42}$ are dimensionless functions listed in the appendix. There they are expressed in terms of hypergeometric functions and double series in the dimensionless variables $x$ and $y = \kappa_0 D/2$.

The resulting structure function for total differential image motion is illustrated in Fig. 2, along with the Kolmogorov case for comparison.

### 2.3 Asymptotic limits

As can be seen from Fig. 2, the mean square differential image motion reaches a limiting value when $\alpha z \gg L_0$. Once the separation between beams exceeds the outer scale, the image motion of the two stars becomes uncorrelated. Further increasing the separation has no effect on the variance.

For Kolmogorov turbulence, the limiting value can be found by taking the limit $z\alpha \to \infty$ in Eqn (5). We obtain

$$\mathcal{D}(\infty) = 2\mathcal{D}_{\pm}(\infty) = \frac{41.68\,J}{2^{8/3}\pi^{1/2}D^{1/3}}\Gamma\left[\begin{array}{c}1/6, 4/3\\17/6, 11/6\end{array}\right]$$
$$= \frac{11.35\,J}{D^{1/3}}, \qquad (10)$$

where

$$J = \int_0^{\infty} C_n^2(z)dz \qquad (11)$$

is the turbulence integral. The notation

$$\Gamma\left[\begin{array}{c}a_1, a_2, \ldots, a_p\\b_1, b_2, \ldots, b_q\end{array}\right] = \frac{\Gamma(a_1)\Gamma(a_2)\cdots\Gamma(a_p)}{\Gamma(b_1)\Gamma(b_2)\cdots\Gamma(b_q)} \qquad (12)$$

is used to represent ratios of products of gamma functions $\Gamma(x)$.

For the von Kármán case, we take the limit $\alpha z \to \infty$ in the series developed in Appendix A2. This gives the result

$$\mathcal{D} = \frac{41.68\,J}{D^{1/3}}\left\{\frac{1}{2^{8/3}\pi^{1/2}}\Gamma\left[\begin{array}{c}1/6, 4/3\\11/6, 17/6\end{array}\right]{}_1F_2\left(\frac{4}{3};\frac{5}{6},\frac{17}{6};\mu^2\right)\right.$$
$$\left. - \frac{9(2\mu)^{1/3}}{40}{}_1F_2\left(\frac{3}{2};\frac{7}{6},3;\mu^2\right)\right\}. \qquad (13)$$

where $\mu = \pi D/L_0$, and ${}_1F_2$ is a generalized hypergeometric function. In terms of numerical values this becomes

$$\mathcal{D} = \frac{11.35\,J}{D^{1/3}}\left[{}_1F_2\left(\frac{4}{3};\frac{5}{6},\frac{17}{6};\mu^2\right)\right.$$
$$\left. - 1.0412\,\mu^{1/3}\,{}_1F_2\left(\frac{3}{2};\frac{7}{6},3;\mu^2\right)\right]. \qquad (14)$$

Asymptotic values for the differential image motion, as a function of $L_0/D$ are shown in Fig. 3. We see that as the outer scale increases, so does the limiting value.

### 2.4 Small separations

As $\alpha z \to 0$, clearly $\mathcal{D}(\alpha) \to 0$. The form of $\mathcal{D}(\alpha)$ at small values can be found by keeping only the first-order terms in the series developed in Appendix A.1. (For small separations the outer scale is unimportant so we can use the Kolmogorov equations.) We obtain

$$\mathcal{D}_+ = \frac{20.84\,\alpha^2}{4 \cdot 2^{8/3}\pi^{1/2}D^{7/3}}\Gamma\left[\begin{array}{c}1/6,1/3\\11/6,5/6\end{array}\right]\int_0^\infty C_n^2(z)z^2dz, \quad (15)$$

$$= \frac{6.501\,\alpha^2}{D^{7/3}}\int_0^\infty C_n^2(z)z^2dz, \quad (16)$$

$$\mathcal{D}_- = \frac{1}{3}\mathcal{D}_+. \quad (17)$$

We see that for small angles, $\mathcal{D}_+(\alpha) = 3\mathcal{D}_-(\alpha) \propto \alpha^2$. The longitudinal variance is three times the transverse variance. We also see that, for small angles, the structure functions are proportional to the integral of the *square* of the distance $z$ to the turbulence, weighted by the turbulence strength $C_n^2(z)$.

### 2.5 Power spectral density

In order to measure the differential image motion, one acquires images that have a finite exposure time. A longer exposure time allows fainter stars to be detected. But at the same time, temporal averaging of the atmospheric fluctuations reduces the differential motion by an amount that depends on the wind speed and turbulence profile of the atmosphere. To quantify this, we must first determine the temporal power spectral density of the differential tilt component of the wavefront phase.

Following (Sasiela 2007 Section 2.4), the temporal power spectrum $S_\pm(\omega)$, where $\omega$ is the angular frequency in radians $s^{-1}$, is related to the structure function by

$$\mathcal{D}_\pm = \frac{1}{\pi}\int_0^\infty S_\pm(\omega)d\omega. \quad (18)$$

Under the assumption of frozen flow, the temporal spectrum can be found from the spatial spectrum by a change of variables. If $\boldsymbol{v}$ is the wind velocity, $\omega$ is related to the spatial frequency vector $\boldsymbol{\kappa}$ by $\omega = \boldsymbol{\kappa} \cdot \boldsymbol{v}$.

Defining $q = \kappa v/\omega$, it follows from Eqns. (2), (3), (4) and (18) that the power spectral density of total differential tilt $S = S_+ + S_-$ is

$$S = \frac{83.36\,\omega}{D^2}\int_0^\infty dz\frac{C_n^2}{v^2}\int_0^\infty qdq\frac{U(q-1)}{\sqrt{q^2-1}}\cos^2\left(\frac{\omega^2q^2z}{2kv^2}\right)$$
$$\times\left\{1 - \cos\left[\frac{\omega\alpha z}{v}(\cos\phi_\alpha + \sqrt{q^2-1}\sin\phi_\alpha)\right]\right\}$$
$$\times f\left(\frac{\omega q}{v}\right)J_1^2\left(\frac{\omega qD}{2v}\right). \quad (19)$$

Here $U$ is the Heaviside step function, $v$ is the wind speed and $\phi_\alpha$ is the angle between the direction of the wind and the vector connecting the two stars, projected on a horizontal plane. In general, both $v$ and $\phi_\alpha$ are functions of $z$.

Since we do not generally know the wind direction, and there are many possible directions between pairs of stars, it is more useful to consider the limit of large separations. In this limit the variance of differential tilt is twice the tilt

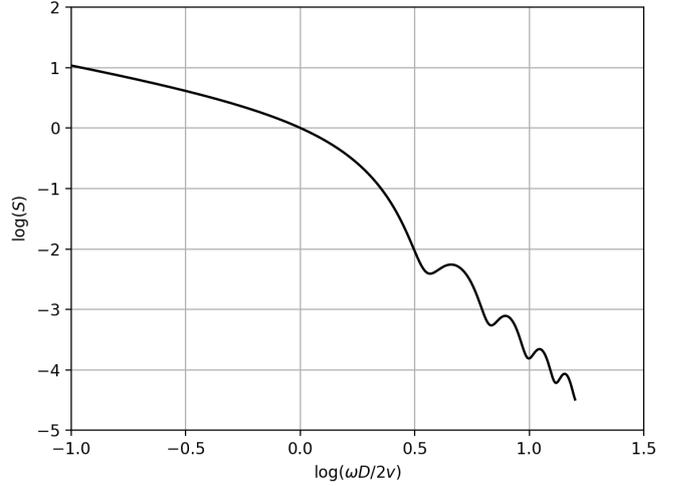

**Figure 4.** Power spectral density of total angular G-tilt, as a function of $x = \omega D/2v$, for Kolmogorov turbulence. It is assumed that velocity is constant along the path. The curve is normalized so that $S(1) = 1$. At low frequencies, $S \propto \omega^{-2/3}$, and at high frequencies, $S \propto \omega^{-11/3}$

variance of a single star and does not depend on orientation. Ignoring diffraction, Eqn (19) then reduces to

$$S(x) = \frac{83.36\,\omega}{D^2}\int_0^\infty dz\frac{C_n^2}{v^2}\int_0^\infty qdq\frac{U(q-1)}{\sqrt{q^2-1}}f\left(\frac{\omega q}{v}\right)$$
$$\times J_1^2\left(\frac{\omega qD}{2v}\right). \quad (20)$$

This integral is evaluated in Appendix A.3 for the case of Kolmogorov turbulence. The result is

$$S = 7.815\,D^{2/3}\int_0^\infty dz\frac{C_n^2}{v}\left\{\frac{1}{\pi^{1/2}}\Gamma\left[\begin{array}{c}-1/3,11/6\\10/3,7/3\end{array}\right]\right.$$
$$\times\,_2F_3\left[\frac{11}{6},\frac{1}{2};\frac{10}{3},\frac{7}{3},\frac{4}{3};-x^2\right] + \frac{\pi^{1/2}}{4}x^{-2/3}\Gamma\left[\begin{array}{c}1/3\\5/6\end{array}\right]$$
$$\left.\times\,_2F_3\left[\frac{3}{2},\frac{1}{6};3,2,\frac{2}{3};-x^2\right]\right\}, \quad (21)$$

where $x = \omega D/2v$ is a dimensionless frequency variable. The spectrum has asymptotic slopes of $\omega^{-2/3}$ for $\omega \ll 2v/D$ and $\omega^{-11/3}$ for $\omega \gg 2v/D$.

The main difference between this spectrum and that obtained by Sasiela (2007 Section 6.7) for the case of Z-tilt, is that at high frequencies the power falls less rapidly for G-tilt, by a factor of $\omega^2$. This is a consequence of the different asymptotic dependence of the G-tilt and Z-tilt filter functions. For von Kármán turbulence, a second break is introduced that flattens the spectrum at low frequencies, $\omega \ll 2\pi v/L_0$ (see Sasiela 2007 Section 11.6).

### 2.6 Effect of a finite exposure time

We can now assess the impact of a finite exposure time. Following Greenwood (1977), the reduction in variance $\mathcal{D}$ can be found by integrating the product of the temporal turbulence spectrum $S(\omega)$ with the rejection transfer function

The variances of the longitudinal and transverse image motion are given by Equations (6) and (9) for Kolmogorov and von Kármán turbulence respectively. In this case, $x$ is not a function of $z$, so the integration over $z$ just gives the turbulence integral $J$. As is well known, these instruments are not sensitive to the distance to the turbulence.

As a check, we can compare our results with existing theory for the DIMM (Sarazin & Roddier 1990), by evaluating the leading terms in the series expansion of Eqn. (6). Referring to the equations in Appendix A.1, and noting that for the DIMM, $x > 1$, we find the exact result for the Kolmogorov case to be

$$\mathcal{D}_\pm = \frac{20.84\Gamma(1/6)J}{2^{14/3}\Gamma(5/6)D^{1/3}} \Bigg\{ \frac{8\Gamma(1/3)}{5\pi^{1/2}\Gamma(17/6)} \\ - x^{-1/3}\bigg[{}_3F_2\left(\frac{3}{2},\frac{1}{6},\frac{1}{6};3,2;x^{-2}\right) \\ \mp \frac{1}{5}\,{}_3F_2\left(\frac{3}{2},\frac{7}{6},-\frac{5}{6};3,2;x^{-2}\right)\bigg]\Bigg\}. \quad (25)$$

If we approximate this by keeping only the first term in each hypergeometric series, and make use of Eqn. (27), we obtain

$$\mathcal{D}_\pm = 2\lambda^2 r_0^{-5/3}[0.1707 D^{-1/3} - 0.1217(1\mp 0.2)r^{-1/3}], \quad (26)$$

Here $r_0$ is the Fried parameter, related to the turbulence integral, and to $\varepsilon$, the full-width at half maximum of stellar images, by the standard relations (Roddier 1981)

$$r_0^{-5/3} = 16.70\,\lambda^{-2} J, \quad (27)$$
$$\varepsilon = 0.976\,\lambda/r_0. \quad (28)$$

Eqn. (26) can be compared with Eqns. (13) and (14) in Sarazin & Roddier (1990). The small differences in the numerical values arise from our use of the equations for G-tilt rather than Z-tilt. For a discussion of the distinction see Sivaramakrishnan et al. (1995) and Chapter 3 of Sasiela (2007).

## 3 METHOD

The nonlinear shape of the relation between differential image motion and angular separation provides an opportunity to estimate the $C_n^2$ profile. At a given angular separation $\alpha$, the physical separation $\alpha z$ is what determines the position on the curves shown in Fig. 1. For a fixed angle $\alpha$, turbulence that is close to the telescope will result in a nearly linear relation between RMS differential image motion and angular separation, whereas turbulence that is far from the telescope will give a strongly nonlinear relation. So if measurements are made for many separations, the shape of the observed $\mathcal{D}_\pm(\alpha)$ relations can be used to infer the vertical distribution of turbulence. In order to resolve high-altitude turbulence, angular separations of a few arcsec are needed, while to resolve turbulence close to the ground, angular separations of several degrees are needed.

### 3.1 Response functions

In order to quantify this idea, observe that Eqn (5) can be written in the form

$$\mathcal{D}_\pm(\alpha) = \int_0^\infty C_n^2(z) W_\pm(\alpha,z) dz, \quad (29)$$

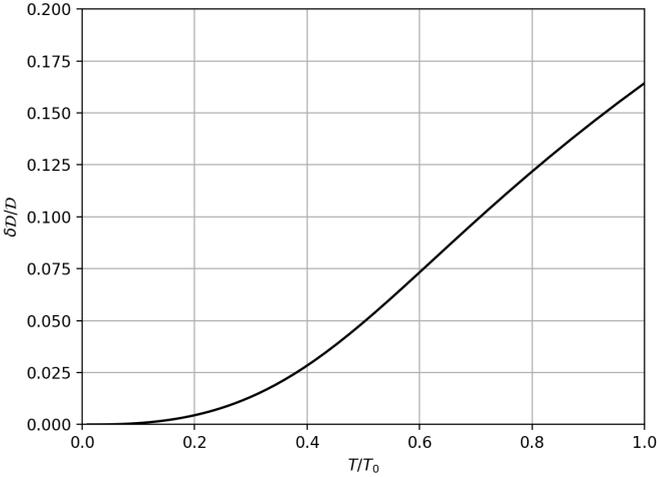

Relative error in the structure constant of differential tilt due to temporal averaging with exposure time $T$. The error depends on the dimensionless ratio $T/T_0$, where $T_0 = \pi D/v_{8/3}$ is a characteristic time determined by the telescope aperture diameter and mean wind speed $v_{8/3}$ defined by Eqn. (24).

$|1 - H(\omega/\omega_c)|^2$, where $H$ is the Fourier transform of the impulse response and $\omega_c$ is a characteristic frequency of the system,

$$\delta \mathcal{D} = \frac{1}{\pi} \int_0^\infty |1 - H(\omega/\omega_c)|^2 \, S(\omega) d\omega. \quad (22)$$

A finite exposure time $T$ is equivalent to temporal averaging of the signal with a boxcar filter of width $T$. The response function is the Fourier transform of the filter, namely

$$H = \mathrm{sinc}(\omega/\omega_c), \quad (23)$$

where $\omega_c = 2\pi/T$.

The filter cuts off low frequencies, so it is sufficient to use the Kolmogorov result for the power spectrum, Eqn. (21). The integral is dominated by the high frequency behavior, which is $S(\omega) \propto C_n^2 v^{8/3} \omega^{-11/3}$. For an extended turbulence distribution, we see that the error depends on the weighted mean velocity

$$v_{8/3} \equiv \left[\frac{1}{J}\int_0^\infty C_n^2(z) v(z)^{8/3} dz\right]^{3/8}. \quad (24)$$

Fig. 2.6 shows the ratio $\delta\mathcal{D}/\mathcal{D}$, computed from Eqns. (22) and (18), as a function of the ratio of exposure time $T$ to the characteristic time $T_0 = \pi D/v_{8/3}$. We see that the error introduced by temporal averaging is less than $\sim 1\%$ for $T/T_0 < 0.2$ and is negligible for $T/T_0 < 0.1$.

### 2.7 Parallel beams

The analytical results obtained here can also be applied to related instruments such as the DIMM and the Generalized Seeing Monitor. In these, a single star is observed through two or more apertures and the differential motion of the star images formed by the individual apertures is measured. For any two apertures, this corresponds to the special case of constant $x = r/D$, where $r$ is the separation between the centres of the apertures and $D$ is the aperture diameter.

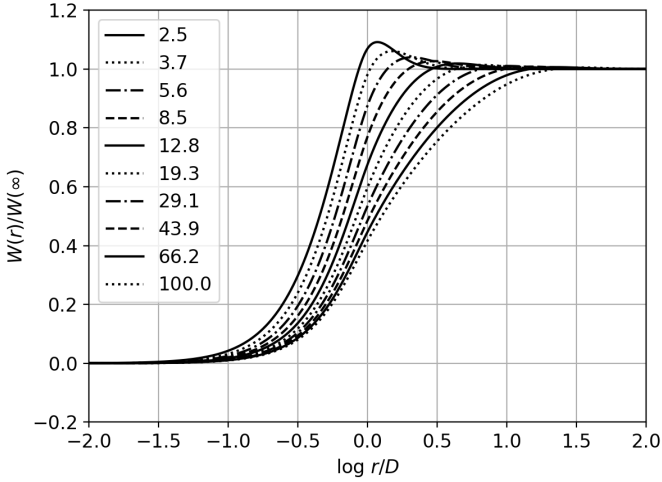

**Figure 5.** Weight functions for total variance, produced by a thin turbulent layer, for the case of von Kármán turbulence. The parameter $r = \alpha z$ corresponds to the physical distance between the centres of the two beams in the turbulent layer. The curves correspond to different values of $L_0/D$.

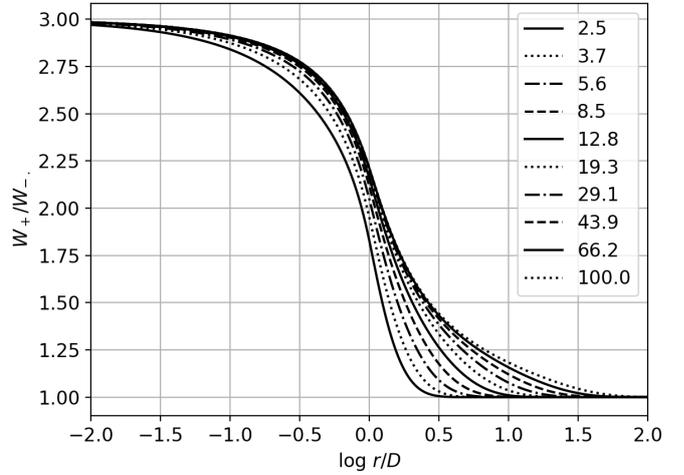

**Figure 6.** The ratio of longitudinal to transverse weight functions, for von Kármán turbulence is plotted vs. $r/D$. For a single thin turbulent layer, this is equal to the ratio of longitudinal to transverse variance. The curves correspond to different values of $L_0/D$.

where $W_\pm(\alpha, z)$ is the function

$$W_\pm(\alpha, z) = \frac{20.84}{D^2} \int_0^\infty d\kappa\, \kappa f(\kappa) J_1^2(\kappa D/2) \times [1 - J_0(\kappa z \alpha) \pm J_2(\kappa z \alpha)]. \quad (30)$$

Equation (29) is a Fredholm integral equation relating the functions $\mathcal{D}_\pm(\alpha)$ and $C_n^2(z)$. In principle, one can invert this equation to find $C_n^2(z)$ from measurements of $\mathcal{D}_\pm(\alpha)$.

Examples of the response functions $W_\pm(\alpha, z)$ are shown in Fig. 5, for several values of $L_0/D$. It can be seen that the differential image motion depends primarily on the ratio of the $r = \alpha z$, the physical separation in the turbulence layer, to the beam diameter $D$. However, the shape of the function depends on the ratio $L_0/D$.

### 3.2 Ratio of longitudinal and transverse variance

It is interesting to examine the ratio of the longitudinal and transverse motions. This ratio is shown in Fig. 6. It depends primarily on $r/D$ and is not strongly affected by $L_0$. We have seen that as $\alpha \to 0$, the ratio approaches three. When $r = D$ it is about two. At large separations the ratio is unity, as the beams are no longer correlated.

### 3.3 Inversion technique

Direct inversion of Eq (29) is not advisable due to measurement noise and statistical fluctuations. Rather, an optimization procedure is preferred, in which one models the $C_n^2$ profile using a finite set of parameters $m_i$. One then adjusts the parameter values to minimize the mean-square difference between the observed values of $\mathcal{D}_\pm(\alpha_j)$ and the variances predicted by Eqn (5), for a series of measurements of stars having a range of separations $\alpha_j$.

We employ a Markov-Chain Monte-Carlo (MCMC) technique to sample the posterior probability $P(m|d,p)$ of the model $m$ given the data $d$ and prior knowledge $p$. By Bayes' theorem, this is given by the product of the probability distribution of the priors and the probability of the data given the model (the likelihood function),

$$P(m|d,p) = P(p)P(d|m) \quad (31)$$

The prior information could include the expected distribution functions of the model parameters. A sensible choice is to constrain the magnitude of $C_n^2(z)$ to have a log-normal distribution at any altitude.

The likelihood function has the form

$$P(d|m,p) = \exp\left[-\frac{1}{2} \boldsymbol{d}^T \boldsymbol{C}^{-1} \boldsymbol{d}\right]. \quad (32)$$

Here $\boldsymbol{d}$ is a column vector whose values are the differences between the observed longitudinal and transverse variances and those predicted by the model. The matrix $C^{-1}$ is the inverse of the covariance matrix

$$C_{ij} = \langle \delta_i \delta_j \rangle - \langle \delta_i \rangle \langle \delta_j \rangle \quad (33)$$

where the $\delta_i$ are the differences in the angular position coordinates.

Once the best-fit parameters for the model have been determined, the model's $C_n^2(h)$ function can be integrated above any desired height $h$. From that, $r_0$ and $\varepsilon$ can be found, as a function of telescope height, from Eqns. (27) and (28).

The MCMC technique also provides estimates of the confidence intervals for all model parameters. This is done by allowing the chain to sample the posterior probability distribution and then computing the resulting standard deviations for each parameter. As the entire n-dimensional distribution is sampled, the resulting 1-sigma confidence intervals are accurate even if there are correlations between model parameters.

# 4 SIMULATIONS

In order to check the viability and performance of the method, several numerical simulations were performed. In each, a test $C_n^2$ profile was chosen and used to generate simulated structure functions. A total of 30 stars were then placed randomly within the same angular field as that of the AST3 telescope. For every pair of stars, a point in the longitudinal and transverse structure functions was generated by random draws from a gaussian distribution having a variance computed from the $C_n^2$ profile using Eqn (9) and the assumed value of $L_0$. The MCMC inversion method was then applied to these data to give an estimated $C_n^2$ profile and estimates of the total seeing, free-atmosphere (FA) seeing, GL thickness, and outer scale.

Simulations are shown in Fig. 7 for rectangular, exponential, and gamma test profiles (defined in Section 5.3). The model that was used in the MCMC analysis has 8 free parameters, consisting of the values of $C_n^2$ at 6 linearly-spaced heights in the GL, spanning the range $3 - 60$ m, the FA seeing, and the outer scale $L_0$. In all cases, the simulated $C_n^2$ profile has total seeing of 1.40 arcsec, FA seeing of 0.30 arcsec, a $C_n^2$-weighted mean GL height $h_G$ of 20 m and an outer scale of 20 m. Recovered values of these parameters are indicated in the figure sub-panels, along with standard errors.

We see that in all cases the recovered parameter values agree with the simulated values, within the estimated errors. Also, the recovered $C_n^2$ profiles are consistent with the simulated profiles.

# 5 OBSERVATIONS

## 5.1 Data

The method was tested using two sets of data taken with the second AST3 telescope at Dome A. It is a modified Schmidt telescope having a 0.5-m aperture and a 2.9° field of view. The telescope is equipped with a frame-transfer CCD camera having 10K×5K active pixels, each 1.0 arcsec square. The typical full-width at half maximum intensity of stellar images is about 4 arcsec (Ma et al. 2018), so the images are well-sampled even with this large pixel size. Two sequences of 30 consecutive exposures of 10 ms duration were used. The time interval between each exposure in the sequence was approximately four seconds, corresponding to the read time of the CCD. As this is much longer than the characteristic timescale $T_0$, discussed in Section 6, the images can be regarded as statistically independent. One sequence was taken in July and the second in August of 2017. After pre-processing, centroid positions of all bright stars (signal-to-noise ratio > 20) were determined using SExtractor (Bertin & Arnouts 1996).

For each star $i = 1, \ldots, N$, the mean angular positions $\bar{x}_i$ and $\bar{y}_i$ were computed along with the displacements $dx_i = x_i - \bar{x}_i$ and $dy_i = y_i - \bar{y}_i$ in each frame. Here the bar indicates the mean of all frames. Next, all possible position vectors $\boldsymbol{r}_{ij} = (\bar{x}_i - \bar{x}_j, \bar{y}_i - \bar{y}_j)$ connecting pairs of stars are computed, and for each pair the differences $\delta x_{ij} = dx_i - dx_j$ and $\delta y_{ij} = dy_i - dy_j$ were computed. These are converted to longitudinal (parallel to $\boldsymbol{r}$) and transverse components by the rotation

$$\delta_{+ij} = \delta x_{ij} \cos\theta_{ij} + \delta y_{ij} \sin\theta_{ij} \tag{34}$$
$$\delta_{-ij} = -\delta x_{ij} \sin\theta_{ij} + \delta y_{ij} \cos\theta_{ij}, \tag{35}$$

where $\theta_{ij} = \arctan(y_{ij}/x_{ij})$ is the polar angle of $\boldsymbol{r}_{ij}$. Finally, the variances $\mathcal{D}_\pm(r) = \operatorname{Var}\delta_\pm$ were computed for every separation.

## 5.2 Field rotation

We discovered that some data were affected by field rotation, which increases the variance of the tangential component of image motion. The AST3 telescopes have equatorial mounts, but any misalignment of the polar axis will introduce field rotation. The observed rotation is very small, on the order of a few microradians over the duration of an image sequence. However that is enough to significantly affect the data given the large field of view. This was verified by plotting the transverse displacements $\delta t_{ij}$ as a function of frame number, where a clear linear trend was seen. The data analysis program was then modified to fit a first-order polynomial to the transverse displacements, and to then subtract this polynomial in order to remove the effect of field rotation.

## 5.3 Models

In the analysis that we employed, simple models of the $C_n^2$ profile were constructed, and the predicted structure functions were then compared with the data in order to find the best model parameters. Two different types of models were employed. So-called "gamma" models have the form

$$C_n^2(h) = A h^n \exp(-h/h_0) + B. \tag{36}$$

Here $h$ is the height above the telescope and $A, B$ and $h_0$ are free parameters. The parameter $B$ represents the value of $C_n^2$ in the FA, which is assumed to be a constant up to some maximum height $h_\mathrm{FA}$ and zero above that. The parameter $n$, not necessarily an integer, controls the mean height of the turbulence in the GL. We do not distinguish between GL and *surface layer* (SL) and use the term GL to describe all low-level turbulence. The choice $n = 0$ results in an exponential GL. Larger values increase the height of the maximum turbulence.

The turbulence integral for these models is

$$J = \int_0^{h_\mathrm{FA}} C_n^2 dh = A h_0^n \Gamma(n+1) + J_\mathrm{FA}, \tag{37}$$

where $J_\mathrm{FA} = B\, h_\mathrm{FA}$ is the FA turbulence integral.

"Discrete" models employ as model parameters the values of $\ln(C_n^2)$ at $n-2$ heights in the atmosphere. The heights are logarithmically spaced and chosen to span the GL. To these are added two more free parameters: the FA turbulence integral $J_\mathrm{FA}$ and the outer scale $L_0$. Values of $\ln(C_n^2)$ at other heights are determined by linear interpolation.

"Linear" models are the same as discrete models except the heights are linearly spaced.

For all models, the total seeing $\varepsilon$ and FA seeing $\varepsilon_\mathrm{FA}$ are obtained from the respective turbulence integrals via Eqns. (27) and (28), which give

$$\varepsilon = 5.2853 \lambda^{-1/5} J^{3/5}. \tag{38}$$

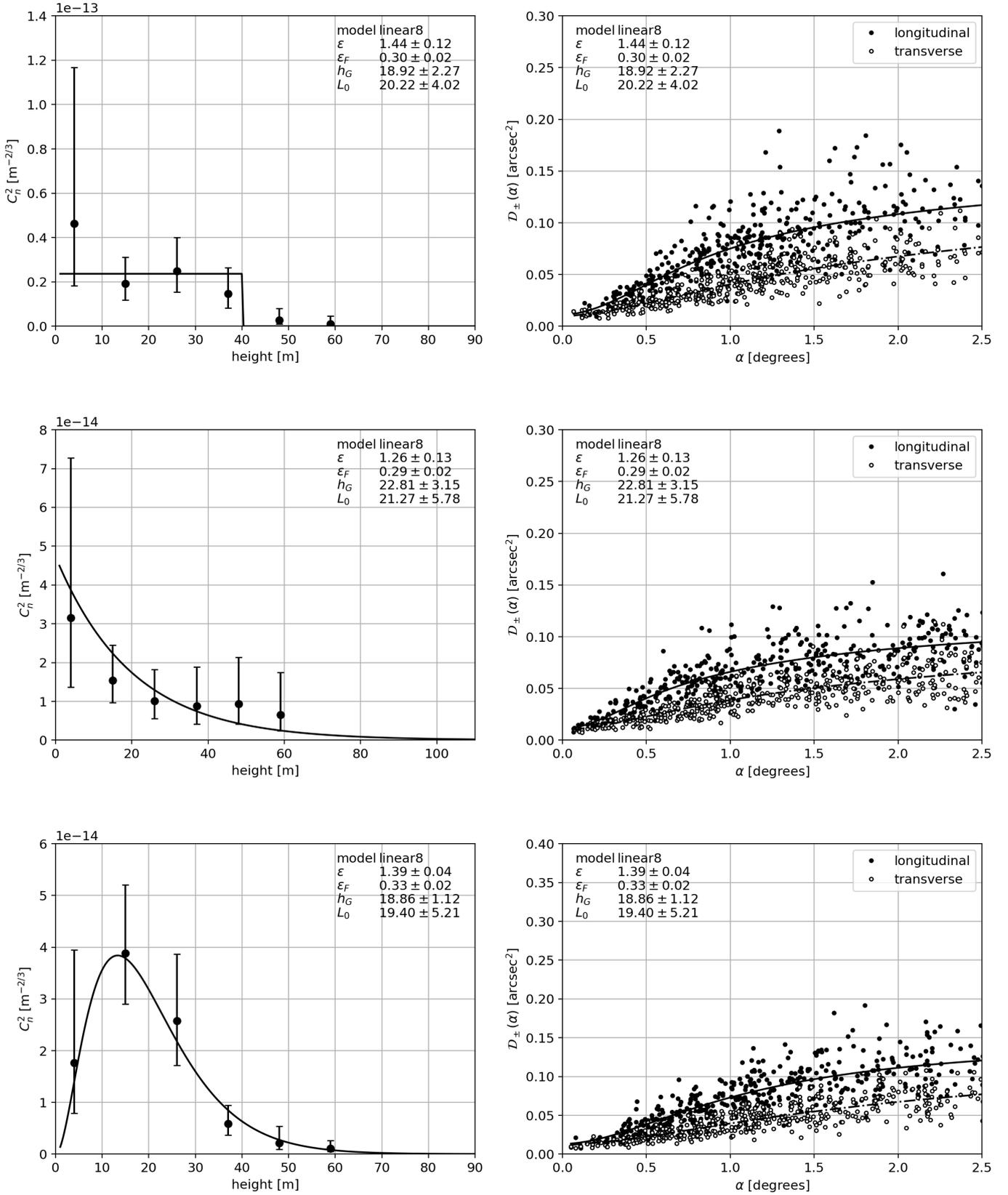

**Figure 7.** Left: Simulated (solid line) and recovered (dots) $C_n^2$ profiles. Right: Structure functions computed from the $C_n^2$ model (solid) and simulated data (dots). Bars indicate standard errors of the recovered $C_n^2$ values at the model heights.

## 6 RESULTS & DISCUSSION

Examples of the data and best-fit models are shown in Figs. 8 and 9. A linear model having 8 free parameters was employed. The heights shown are above the telescope. One should add 4 m to obtain the height above the ground. The parameter $h_{\rm GL}$ is the mean $C_n^2$-weighted height of the ground layer. Other models were also tested, including discrete models with up to 10 free parameters, and gamma models. The results for the integrated quantities, shown in the figures, were essentially identical, within the statistical errors.

It is immediately evident that the ground layer was very different when the two data sets were obtained. The July data show a very thin and relatively weak GL, while August data show a GL that is much thicker with greater total turbulence. This is consistent with SNODAR results at Dome A, which show strong seasonal variations in the GL thickness, and a median thickness of 13.9 m Bonner et al. (2010). (The definition of thickness used by Bonner et al. (2010), the height at which $C_T^2$ falls to 1% of its initial value, is roughly two times larger than the mean height, $h_G$).

For these data sets, a FA seeing in the range $0.29 - 0.33 \pm 0.02$ arcsec was found. This can be compared with the median value of 0.36 arcsec found for Dome C Aristidi et al. (2009). However, we emphasize that these two examples are only illustrative. They do not represent the average seeing and turbulence structure, and should not be used to describe the Dome A site. One cannot obtain meaningful statistical results from just these two very-small data sets. A study of a much-larger data set is currently in progress and will be the subject of a future paper.

What is the impact of the 10 ms exposure time used for these data? At Dome A, most of the turbulence integral is contributed by the GL. The wind velocity in this layer is typically quite small, having a yearly average of $3 - 5$ m s$^{-1}$ (Hu et al. 2014, 2019). For the AST3 telescopes, and a wind speed of 5 m s$^{-1}$, the characteristic time is $T_0 = 0.3$ s. From Fig. 2.6 we see that the exposure time of 10 ms ($T/T_0 \simeq 0.03$) results in negligible error. In fact, it should be possible to increase the exposure time without introducing a significant error. That would allow fainter stars to be detected, increasing the number of baselines and allowing smaller separations to be sampled, improving the signal-to-noise ratio and the sensitivity to higher-altitude turbulence. The optimal exposure time will depend on the turbulence-weighted mean wind speed at the site, including the contribution of the free atmosphere.

Another question that arises is the number of frames $N_f$ that should be acquired for each turbulence profile measurement. There are several considerations. More frames result in better statistical averaging and therefore lower errors in the measurement of the structure functions. These errors would be expected to decrease in proportion to $N_f^{-1/2}$. Of course more frames result in a lower temporal resolution for the measurement of variations in the turbulence profile. The time required per frame is determined by the readout time of the camera, which may be as large as several seconds. A reasonable compromise for the AST3 telescopes is a profile measurement every two minutes.

A further consideration is the length of time needed to obtain a reasonable estimate of the outer scale. Here, the relevant time scale is $L_0/v$, where $v$ is the typical wind speed. Ideally, one should sample many statistical realizations of the turbulence when estimating the outer scale. At Dome A, $L_0/v \sim 4$ s, so a two-minute observation will sample $\sim 30$ independent realizations. This results in a statistical error of about 20%, which is consistent with the estimated errors shown in Figs. (8) and (9).

## 7 CONCLUSIONS

We have described a new technique for the measurement of atmospheric turbulence, the Atmospheric Distortion Monitor, and have provided examples of its application. The MTM has the ability to simultaneously estimate low-resolution turbulence profiles and the outer scale. The height within which the turbulence profile can be resolved is limited by the smallest angular separation $\alpha_{\rm min}$ of the observed stars. As the differential motion begins to saturate when the linear separation between beams in the turbulent layer exceeds the telescope diameter, the upper limit is on the order of $z_{\rm max} \simeq D/\alpha_{\rm min}$. For the examples presented, $\alpha_{\rm min} \simeq 0.1°$ which corresponds to a height of $\sim 300$ m. By observing galactic clusters containing close pairs of stars, this limit could be extended perhaps to several km. However, it should be noted that the structure function decreases as the separation decreases, so will be more difficult to measure. The AST3 telescopes should in principle be able to measure stars as faint as 10th magnitude in a 10 ms exposure, and fainter with longer exposure times. We hope to try this in the near future.

The method described here has similarities with the technique of measuring correlations in fluctuations in images of the lunar limb (Maire et al. 2007; Ziad et al. 2010a,b). But it differs in several respects. Our technique employs both components of angular motion, does not require that the Moon be visible, is not limited by the finite angular size of the Moon, and should in principle be able to provide a larger number of data points, hence better resolution of the structure functions.

The MTM technique is also similar to that of the S-DIMM (Scharmer & van Werkhoven 2010), in which turbulence profiles are obtained from the structure function of differential angular motion of solar granulation. The two techniques are complimentary as one can be used in the day time and the other at night. However, there are other differences. In the S-DIMM technique, it is necessary to average over a finite area on the solar disk in order to determine the displacements of areas of granulation. This, and the finite size of the Sun, limits the dynamic range.

We are grateful to an anonymous referee whose suggestions helped improve the manuscript. PH acknowledges financial support from the Natural Sciences and Engineering Research Council of Canada and thanks NAOC for hospitality during a sabbatical visit made possible by support from the Chinese Academy of Sciences (CAS President's International Fellowship Initiative, 2017VMA0013). We also acknowledge support from the National Natural Science Foundation of China under grant numbers 11733007 and 11673037.

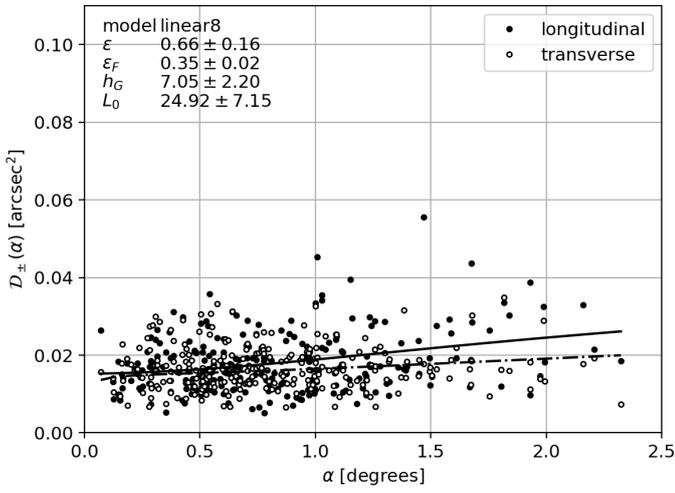 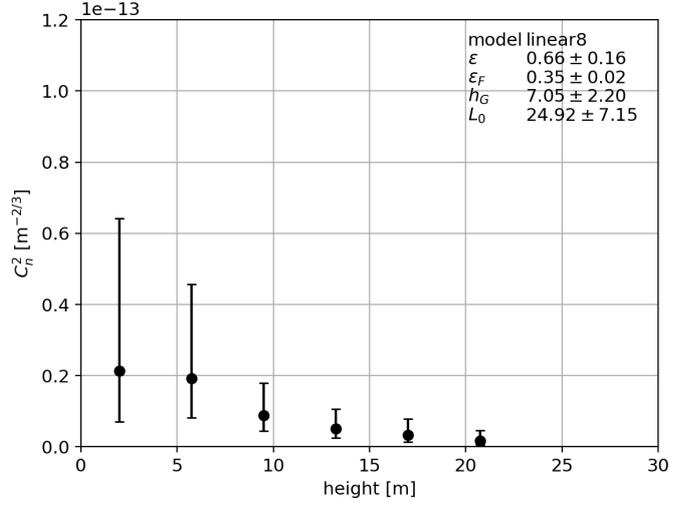

**Figure 8.** Left: Observed and modelled (solid lines) structure functions for the July 2010 data set. Right: $C_n^2$ profile derived from the best-fit model.

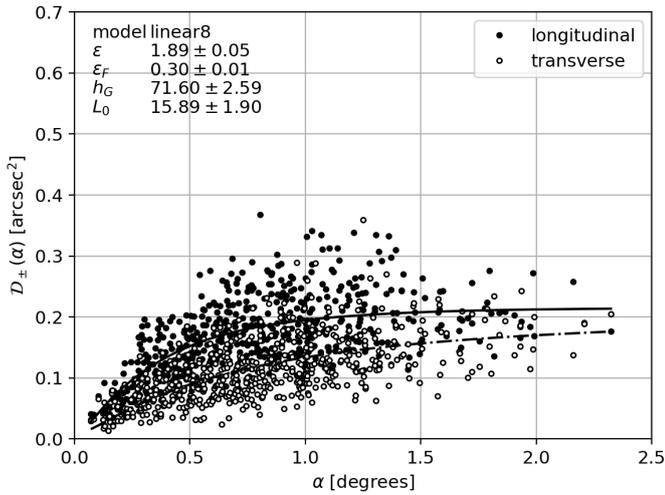 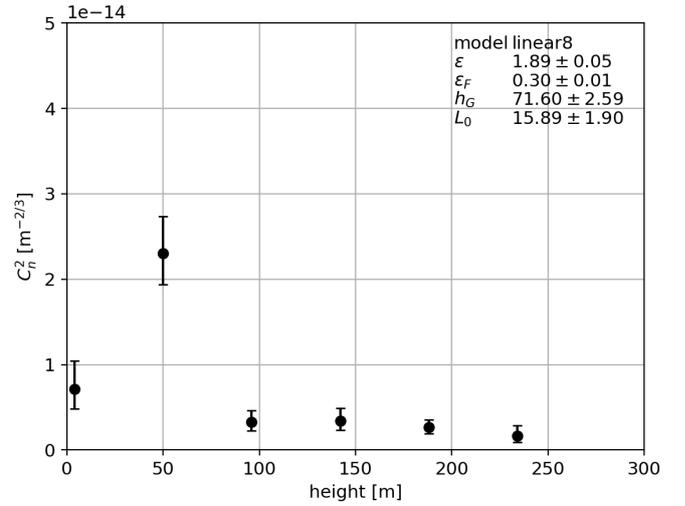

**Figure 9.** Left: Observed and modelled (solid lines) structure functions for the August 2010 data set. Right: $C_n^2$ profile derived from the best-fit model.

# Appendices

## A EVALUATION OF INTEGRALS

### A.1 Kolmogorov spectrum

We wish to evaluate the integrals

$$I_\pm = D^{-5/3}\int_0^\infty d\kappa\, \kappa^{-8/3}J_1^2\left(\frac{\kappa D}{2}\right)[1 - J_0(\kappa z\alpha) \pm J_2(\kappa z\alpha)], \tag{A.1}$$

The factor $D^{-5/3}$ is inserted to make the integrals dimensionless. This integral can be done using Mellin transforms Sasiela (2007). The two integrals that we need are

$$I_1(u) = D^{-5/3}\int_0^\infty d\kappa\, \kappa^{-8/3}J_1^2(\kappa u), \tag{A.2}$$

$$I_{2\nu}(u,v) = D^{-5/3}\int_0^\infty d\kappa\, \kappa^{-8/3}J_1^2(\kappa u)J_\nu(\kappa v), \tag{A.3}$$

where $\nu = 0, 2$, $u = D/2$ and $v = \alpha z$. The relevant Mellin transforms are

$$J_\nu^2(t) \to \frac{1}{2\sqrt{\pi}}\Gamma\left[\begin{array}{c} s/2 + \nu, 1/2 - s/2 \\ 1 + \nu - s/2, 1 - s/2 \end{array}\right], \quad -2\Re(\nu) < \Re(s) < 1, \tag{A.4}$$

$$J_\nu(t) \to 2^{s-1}\Gamma\left[\begin{array}{c} s/2 + \nu/2 \\ 1 + \nu/2 - s/2 \end{array}\right], \quad -\Re(\nu) < \Re(s) < 3/2. \tag{A.5}$$

The gamma function notation used here represents the product of gamma functions of terms in the upper row divided by the product of gamma functions of the terms in the lower row. The first integral has no free parameters as $u$ can be removed from the integral by the substitution $\kappa u = t$. Using the definition of the Mellin transform,

$$\int_0^\infty \frac{dt}{t}t^a f(bt^c) = \frac{b^{-a/c}}{c}F\left(\frac{a}{c}\right), \quad c > 0, \tag{A.6}$$

where $F(s)$ denotes the Mellin transform of $f(t)$, we obtain

$$I_1(u) = \frac{1}{2^{8/3}\pi^{1/2}}\Gamma\left[\begin{array}{c} 1/6, 4/3 \\ 17/6, 11/6 \end{array}\right] \tag{A.7}$$

The second integral has one free parameter. The substitutions $t = \kappa u$, $x = \alpha z/D = v/2u$ convert this to

$$I_{2\nu}(x) = \frac{1}{2^{5/3}}\int_0^\infty dt\, t^{-8/3}J_1^2(t)J_\nu(2xt), \tag{A.8}$$

Using the general result

$$\int_0^\infty \frac{dt}{t}t^a f(bt^c)g(dt^e) = \frac{b^{-a/c}}{ec}\frac{1}{2\pi i}\int_C ds (d^{1/e}b^{-1/c})^s F\left(\frac{s+a}{c}\right)G\left(-\frac{s}{e}\right), \quad c > 0, \tag{A.9}$$

we obtain,

$$I_{2\nu} = \frac{1}{2^{11/3}\pi^{1/2}}\frac{1}{2\pi i}\int_C ds\, x^s \Gamma\left[\begin{array}{c} s/2 + 1/6, 4/3 - s/2, -s/2 + \nu/2 \\ 17/6 - s/2, 11/6 - s/2, 1 + \nu/2 + s/2 \end{array}\right]. \tag{A.10}$$

The contour C is the vertical line extending from $\eta - i\infty$ to $\eta + i\infty$, where $\eta$ is a real constant chosen to satisfy the conditions on $s$ imposed by the Mellin transforms employed. This integral can be simplified by the substitution $s \to 2s$,

$$I_{2\nu} = \frac{1}{2^{8/3}\pi^{1/2}}\frac{1}{2\pi i}\int_C ds\, x^{2s}\Gamma\left[\begin{array}{c} s + 1/6, 4/3 - s, -s + \nu/2 \\ 17/6 - s, 11/6 - s, 1 + \nu/2 + s \end{array}\right] \tag{A.11}$$

The gamma function will have a simple pole whenever any of the arguments in the numerator are equal to zero or a negative integer. There are three series of poles, corresponding to $s = -1/6 - n, 4/3 + n, \nu/2 + n$ where $n = 0, 1, 2, \ldots$ Each of these choices gives rise to a series, each of which will have a certain domain of convergence. If the domains overlap, we add the series in those regions, otherwise, we choose the appropriate series for each parameter interval.

The three series are

$$S_{1\nu} = \frac{x^{-1/3}}{2^{8/3}\pi^{1/2}} \sum_{n=0}^{\infty} \frac{(-1)^n}{n!} x^{-2n} \Gamma\left[\begin{array}{c} 3/2+n, 1/6+\nu/2+n \\ 3+n, 2+n, 5/6+\nu/2-n \end{array}\right],$$

$$= \frac{x^{-1/3}}{2^{14/3}} \Gamma\left[\begin{array}{c} 1/6+\nu/2 \\ 5/6+\nu/2 \end{array}\right] {}_3F_2\left(\frac{3}{2}, \frac{1}{6}+\frac{\nu}{2}, \frac{1}{6}-\frac{\nu}{2}; 3, 2; x^{-2}\right), \tag{A.12}$$

$$S_{2\nu} = \frac{x^{8/3}}{2^{8/3}\pi^{1/2}} \sum_{n=0}^{\infty} \frac{(-1)^n}{n!} x^{2n} \Gamma\left[\begin{array}{c} 3/2+n, -4/3+\nu/2-n \\ 3/2-n, 1/2-n, 7/3+\nu/2+n \end{array}\right],$$

$$= \frac{x^{8/3}}{2^{8/3}\pi} \Gamma\left[\begin{array}{c} -4/3+\nu/2 \\ 7/3+\nu/2 \end{array}\right] {}_3F_2\left(\frac{3}{2}, -\frac{1}{2}, \frac{1}{2}; \frac{7}{3}-\frac{\nu}{2}, \frac{7}{3}+\frac{\nu}{2}; x^2\right), \tag{A.13}$$

$$S_{3\nu} = \frac{x^{\nu}}{2^{8/3}\pi^{1/2}} \sum_{n=0}^{\infty} \frac{(-1)^n}{n!} x^{2n} \Gamma\left[\begin{array}{c} \nu/2+1/6+n, 4/3-\nu/2-n \\ 17/6-\nu/2-n, 11/6-\nu/2-n, 1+\nu+n \end{array}\right],$$

$$= \frac{x^{\nu}}{2^{8/3}\pi^{1/2}} \Gamma\left[\begin{array}{c} \nu/2+1/6, 4/3-\nu/2 \\ 17/6-\nu/2, 11/6-\nu/2, 1+\nu \end{array}\right] {}_3F_2\left(\frac{\nu}{2}+\frac{1}{6}, \frac{\nu}{2}-\frac{11}{6}, \frac{\nu}{2}-\frac{5}{6}; \frac{\nu}{2}-\frac{1}{3}, 1+\nu; x^2\right). \tag{A.14}$$

Here ${}_3F_2$ denotes a generalized hypergeometric function. The first series converges for $x > 1$, and the second and third converge for $x < 1$. Therefore, the integrals are

$$I_{\pm} = \begin{cases} I_1 - S_{10} \pm S_{12}, & x \geq 1 \\ I_1 - S_{20} \pm S_{22} - S_{30} \pm S_{32}, & x \leq 1 \end{cases} \tag{A.15}$$

### A.2 von Kármán spectrum

In this case, $f(\kappa) = (\kappa^2 + \kappa_0^2)^{-11/6}$, so the two integrals that we need are

$$I_3 = \frac{D^{-5/3}}{\kappa_0^{11/3}} \int_0^{\infty} d\kappa \kappa [1+(\kappa/\kappa_0)^2]^{-11/6} J_1^2(\kappa u), \tag{A.16}$$

$$I_{4\nu} = \frac{D^{-5/3}}{\kappa_0^{11/3}} \int_0^{\infty} d\kappa \kappa [1+(\kappa/\kappa_0)^2]^{-11/6} J_1^2(\kappa u) J_\nu(\kappa v), \tag{A.17}$$

where $\nu = 0, 2$. To simplify these, we make the substitutions $t = \kappa u, y = \kappa_0 u, x = \kappa_0 v/2$ and obtain

$$I_3 = 2^{-5/3} y^{-11/3} \int_0^{\infty} dt \, t [1+(t/y)^2]^{-11/6} J_1^2(t), \tag{A.18}$$

$$I_{4\nu} = 2^{-5/3} y^{-11/3} \int_0^{\infty} dt \, t [1+(t/y)^2]^{-11/6} J_1^2(t) J_\nu(2xt/y). \tag{A.19}$$

For this we need an additional Mellin transform.

$$(1+t^2)^{-p} \to \frac{1}{2} \Gamma\left[\begin{array}{c} s/2, p-s/2 \\ p \end{array}\right], \quad 0 < \Re(s) < 2\Re(p). \tag{A.20}$$

The first integral has just one parameter and can be done using Eqn (A.9). We have

$$\int_0^{\infty} \frac{dt}{t} t^a f(bt^c) g(dt^e) = \frac{b^{-a/c}}{ec} \frac{1}{2\pi i} \int_C ds (d^{1/e} b^{-1/c})^s F\left(\frac{s+a}{c}\right) G\left(-\frac{s}{e}\right), \quad c > 0, \tag{A.21}$$

which gives

$$I_3 = \frac{y^{-5/3}}{2^{8/3}\sqrt{\pi}\,\Gamma(11/6)} \frac{1}{2\pi i} \int_C ds \, y^{2s} \Gamma\left[\begin{array}{c} 5/6-s, 1-s, 1/2+s \\ 2+s \end{array}\right]. \tag{A.22}$$

There are three series of poles, $s = 5/6+n, 1+n, -1/2-n$. These give rise to the series

$$S_4 = \frac{1}{2^{8/3}\sqrt{\pi}\,\Gamma(11/6)} \sum_{n=0}^{\infty} \frac{(-1)^n}{n!} y^{2n} \Gamma\left[\begin{array}{c} 1/6-n, 4/3+n \\ 17/6+n \end{array}\right] = \frac{1}{2^{8/3}\sqrt{\pi}} \Gamma\left[\begin{array}{c} 1/6, 4/3 \\ 11/6, 17/6 \end{array}\right] {}_1F_2(4/3; 5/6, 17/6; y^2), \tag{A.23}$$

$$S_5 = \frac{y^{1/3}}{2^{8/3}\sqrt{\pi}\,\Gamma(11/6)} \sum_{n=0}^{\infty} \frac{(-1)^n}{n!} y^{2n} \Gamma\left[\begin{array}{c} -1/6-n, 3/2+n \\ 3+n \end{array}\right] = -\frac{9y^{1/3}}{2^{8/3} \cdot 5} {}_1F_2(3/2; 7/6, 3; y^2), \tag{A.24}$$

$$S_6 = \frac{y^{-8/3}}{2^{8/3}\sqrt{\pi}\,\Gamma(11/6)} \sum_{n=0}^{\infty} \frac{(-1)^n}{n!} y^{-2n} \Gamma\left[\begin{array}{c} 4/3+n, 3/2+n \\ 3/2-n \end{array}\right] = \frac{y^{-8/3}}{2^{8/3}\sqrt{\pi}} \Gamma\left[\begin{array}{c} 4/3 \\ 11/6 \end{array}\right] {}_3F_0(4/3, 3/2, -1/2; y^{-2}). \tag{A.25}$$

The first two series converge for any value of the argument $y^2$. The third series diverges for any value of its argument. Therefore, the integral is

$$I_3 = \frac{1}{2^{8/3}}\left\{\frac{1}{\sqrt{\pi}}\Gamma\left[\begin{array}{c}1/6, 4/3\\11/6, 17/6\end{array}\right]\,_1F_2(4/3; 5/6, 17/6; y^2) - \frac{9y^{1/3}}{5}\,_1F_2(3/2; 7/6, 3; y^2)\right\}. \tag{A.26}$$

The first few terms in the series are

$$I_3 = \frac{1}{2^{8/3}}\left\{\frac{1}{\sqrt{\pi}}\Gamma\left[\begin{array}{c}1/6, 4/3\\11/6, 17/6\end{array}\right]\left(1+\frac{48}{85}y^2\right) - \frac{9y^{1/3}}{5}\left(1+\frac{3}{7}y^2\right) + \cdots\right\}. \tag{A.27}$$

The second integral has two free parameters. From the Mellin convolution theorem,

$$\int_0^\infty \frac{dt}{t}t^a f(bt)g(ct)h(dt) = \frac{b^{-a}}{(2\pi i)^2}\int_C ds \int_C dt (c/b)^s (d/b)^t F(s+t+a)G(-s)H(-t). \tag{A.28}$$

Using this, the second integral is

$$I_{4\nu} = \frac{y^{-5/3}}{2^{8/3}\pi^{1/2}\Gamma(11/6)}\frac{1}{(2\pi i)^2}\int_C ds \int_C dt\, y^{2s}x^{2t}\Gamma\left[\begin{array}{c}1+s+t, 5/6-s-t, 1-s, 1/2+s, \nu/2-t\\2+s, 1+s, 1+\nu/2+t\end{array}\right]. \tag{A.29}$$

A 2-pole in $s-t$ space occurs when two arguments of the gamma function take zero or negative integer values. Each possible combination gives rise to a double series. There are eight possible combinations which result in the five convergent series

$$S_{7\nu} = \frac{x^{-4}y^{1/3}}{2^{8/3}\sqrt{\pi}\Gamma(11/6)}\sum_{m,n=0}^\infty \frac{(-1)^{m+n}}{m!n!}x^{-2m}(y/x)^{2n}\Gamma\left[\begin{array}{c}11/6+m, 3/2+n, 2+\nu/2+m+n\\3+n, 2+n, -1+\nu/2-m-n\end{array}\right], \tag{A.30}$$

$$S_{8\nu} = \frac{x^{-1/3}y^{1/3}}{2^{8/3}\sqrt{\pi}\Gamma(11/6)}\sum_{m,n=0}^\infty \frac{(-1)^{m+n}}{m!n!}x^{2m}(y/x)^{2n}\Gamma\left[\begin{array}{c}11/6+m, 3/2+n, 1/6+\nu/2-m+n\\3+n, 2+n, 5/6+\nu/2+m-n\end{array}\right], \tag{A.31}$$

$$S_{9\nu} = \frac{x^\nu y^{1/3}}{2^{8/3}\sqrt{\pi}\Gamma(11/6)}\sum_{m,n=0}^\infty \frac{(-1)^{m+n}}{m!n!}x^{2m}y^{2n}\Gamma\left[\begin{array}{c}2+\nu/2+n+m, -1/6-\nu/2-n-m, 3/2+n\\3+n, 2+n, 1+\nu+m\end{array}\right], \tag{A.32}$$

$$S_{10\nu} = \frac{x^{8/3}y^{-8/3}}{2^{8/3}\sqrt{\pi}\Gamma(11/6)}\sum_{m,n=0}^\infty \frac{(-1)^{m+n}}{m!n!}x^{2m}(x/y)^{2n}\Gamma\left[\begin{array}{c}11/6+m, 3/2+n, -4/3+\nu/2-n-m\\3/2-n, 1/2-n, 7/3+\nu/2+n+m\end{array}\right], \tag{A.33}$$

$$S_{11\nu} = \frac{x^\nu y^{-\nu}}{2^{8/3}\sqrt{\pi}\Gamma(11/6)}\sum_{m,n=0}^\infty \frac{(-1)^{m+n}}{m!n!}(x/y)^{2m}y^{2n}\Gamma\left[\begin{array}{c}11/6+n, 1/6+\nu/2+m-n, 4/3-\nu/2-m+n\\17/6-\nu/2-m+n, 11/6-\nu/2-m+n, 1+\nu+m\end{array}\right]. \tag{A.34}$$

$S_7$ is identically zero, as the third gamma function in the denominator is infinite for all values of $m$ and $n$ when $\nu = 0$ or $\nu = 2$. $S_8$ converges if $x > y$. $S_9$ converges for all values of $x$ and $y$. $S_{10}$, $S_{12}$ and $S_{13}$ diverge for all values of $x$ and $y$ and are excluded. $S_{11}$ and $S_{14}$ converge if $x < y$. Therefore, the final result is

$$I_\pm = I_3 - I_{40} \pm I_{42}, \tag{A.35}$$

where

$$I_{4\nu} = \begin{cases} S_{9\nu} + S_{10\nu} + S_{11\nu}, & x < y, \\ S_{8\nu} + S_{9\nu}, & x \geq y. \end{cases} \tag{A.36}$$

It is interesting to consider the limit at large separations, $r \gg L_0$, where the structure function saturates. This corresponds to $x \to \infty$. $I_2$ is not a function of $x$, so is unchanged. The solution given by Eqn (A.36) involves the difference between increasingly-large numbers, so is not suitable. Instead, an asymptotic series is needed. This can be found using the methods described by Sasiela (2007). We take all series for which the power of $x$ is negative, and add the steepest-descent contribution. As we are interested in the limit $x \to \infty$, only the first term need be retained. This gives the following series contributions,

$$S_{7\nu} \to \frac{x^{-4}y^{1/3}}{2^{8/3}\sqrt{\pi}\Gamma(11/6)}\Gamma\left[\begin{array}{c}11/6, 3/2, 2+\nu/2\\3, 2, -1+\nu/2\end{array}\right], \tag{A.37}$$

$$S_{8\nu} \to \frac{x^{-1/3}y^{1/3}}{2^{8/3}\sqrt{\pi}\Gamma(11/6)}\Gamma\left[\begin{array}{c}11/6, 3/2, 1/6+\nu/2\\3, 2, 5/6+\nu/2\end{array}\right], \tag{A.38}$$

The series contribution to the asymptotic solution is therefore

$$I_{3\nu} \to \frac{x^{-1/3}y^{1/3}}{2^{14/3}}\left\{\Gamma\left[\begin{array}{c}2+\nu/2\\-1+\nu/2\end{array}\right]x^{-11/3} + \Gamma\left[\begin{array}{c}1/6+\nu/2\\5/6+\nu/2\end{array}\right]\right\}. \tag{A.39}$$

This vanishes as $x \to \infty$. The saddle point contribution can also be shown to vanish as $x \to \infty$, so we conclude that, in the limit of large separations, $I_\pm = I_3$.

### A.3 Temporal power spectral density

Here we evaluate, in the near-field limit, the power spectral density of image centroid displacement. The problem is very similar to one already solved by Sasiela (2007 Section 11.6), the only difference being our use of the filter function for G-tilt rather than Z-tilt. Substituting the Kolmogorov spectrum into Eqn. (20) and making the substitution $x = D\omega/2v$, we obtain

$$S = \frac{41.7\,\omega^{-8/3}}{D^2} \int_0^\infty dz C_n^2 v^{5/3} \int_0^\infty \frac{dq}{q} q^{-5/3} \frac{U(q-1)}{\sqrt{q^2-1}} J_1^2(qx). \tag{A.40}$$

In order to evaluate this we need the Mellin transforms given in Eqns. (A.4) and also

$$(t^2-1)^{p-1} U(t-1) \to \frac{1}{2}\Gamma(p)\Gamma\left[\begin{array}{c} 1-p-s/2 \\ 1-s/2 \end{array}\right], \quad \Re(p) > 0, \Re(p+s/2) < 1, \tag{A.41}$$

where $U(t)$ is the Heaviside step function. The integral over $q$ can be done using Eqn. (A.9),

$$I \equiv \int_0^\infty \frac{dq}{q} q^{-5/3} \frac{U(q-1)}{\sqrt{q^2-1}} J_1^2(qx),$$

$$= \frac{1}{2\pi i}\int_C ds x^s \frac{1}{2}\Gamma(1/2)\Gamma\left[\begin{array}{c} 1/2-(s-5/3)/2 \\ 1-(s-5/3)/2 \end{array}\right] \frac{1}{2\sqrt{\pi}}\Gamma\left[\begin{array}{c} -s/2+1, 1/2+s/2 \\ 2+s/2, 1+s/2 \end{array}\right],$$

$$= \frac{1}{4}\frac{1}{2\pi i}\int_C ds x^s \Gamma\left[\begin{array}{c} 4/3-s/2, 1-s/2, 1/2+s/2 \\ 11/6-s/2, 2+s/2, 1+s/2 \end{array}\right]. \tag{A.42}$$

The substitution $s \to -2s$ gives

$$I = \frac{1}{4\pi i}\int_C ds x^{-2s}\Gamma\left[\begin{array}{c} 4/3+s, 1+s, 1/2-s \\ 11/6+s, 2-s, 1-s \end{array}\right]. \tag{A.43}$$

There are three sets of poles, corresponding to $s = -4/3 - n$, $s = -1 - n$ and $s = 1/2 + n$. Convergence is determined by the parameter $\Delta$, which is defined as the sum of the coefficients of $s$ in the gamma functions in the numerator minus those in the denominator. In this case, $\Delta = 2$. The contour can therefore be closed in the left half plane, $\Re(s) < 0$, which encloses the first two sets of poles. These give rise to the series

$$S_1 = \frac{x^{8/3}}{2}\sum_{n=0}^\infty \frac{(-1)^n}{n!} x^{2n}\Gamma\left[\begin{array}{c} -1/3-n, 11/6+n \\ 1/2-n, 10/3+n, 7/3+n \end{array}\right], \tag{A.44}$$

$$S_2 = \frac{x^2}{2}\sum_{n=0}^\infty \frac{(-1)^n}{n!} x^{2n}\Gamma\left[\begin{array}{c} 1/3-n, 3/2+n \\ 5/6-n, 3+n, 2+n \end{array}\right]. \tag{A.45}$$

The result is given by the sum of these two convergent series,

$$S = 5.21\,\omega^{-2/3}\int_0^\infty dz\frac{C_n^2}{v^{1/3}}\sum_{n=0}^\infty \frac{(-1)^n}{n!} x^{2n}\left\{x^{2/3}\Gamma\left[\begin{array}{c} -1/3-n, 11/6+n \\ 1/2-n, 10/3+n, 7/3+n \end{array}\right] + \Gamma\left[\begin{array}{c} 1/3-n, 3/2+n \\ 5/6-n, 3+n, 2+n \end{array}\right]\right\},$$

$$= 5.21\,\omega^{-2/3}\int_0^\infty dz\frac{C_n^2}{v^{1/3}}\left\{\frac{1}{\pi^{1/2}}\left(\frac{D\omega}{2v}\right)^{2/3}\Gamma\left[\begin{array}{c} -1/3, 11/6 \\ 10/3, 7/3 \end{array}\right] {}_2F_3\left[\begin{array}{c} 11/6, 1/2; 10/3, 7/3, 4/3; -\left(\frac{D\omega}{2v}\right)^2 \end{array}\right]\right.$$

$$\left. + \frac{\pi^{1/2}}{4}\Gamma\left[\begin{array}{c} 1/3 \\ 5/6 \end{array}\right] {}_2F_3\left[\begin{array}{c} 3/2, 1/6; 3, 2, 2/3; -\left(\frac{D\omega}{2v}\right)^2 \end{array}\right]\right\}. \tag{A.46}$$

At high frequencies, $\omega \gg 2v/D$, this solution requires the near-cancellation of increasingly large terms, so an asymptotic series is required. Following Sasiela (2007 Section 5.2), it can be obtained from the third series, with the addition of the steepest-descent term. The result is

$$I = \frac{1}{2x}\sum_{n=0}^{n_0} \frac{(-1)^n}{n!} x^{-2n}\Gamma\left[\begin{array}{c} 11/6+n, 3/2+n \\ 7/3+n, 3/2-n, 1/2-n \end{array}\right] - \frac{x^{-3/2}}{2\pi^{1/2}}\cos(2x - \pi/4). \tag{A.47}$$

So the asymptotic result, valid for $\omega > 4v/D$, is

$$S = \frac{41.7\,\omega^{-11/3}}{D^3}\int_0^\infty dz C_n^2 v^{8/3}\left\{\sum_{n=0}^{n_0} \frac{(-1)^n}{n!}\left(\frac{2v}{D\omega}\right)^{2n}\Gamma\left[\begin{array}{c} 11/6+n, 3/2+n \\ 7/3+n, 3/2-n, 1/2-n \end{array}\right] - \left(\frac{2v}{\pi D\omega}\right)^{1/2}\cos\left(\frac{D\omega}{v} - \frac{\pi}{4}\right)\right\}. \tag{A.48}$$

Here $n_0$ is the number of poles of this series that are to the left of the saddle point. In practice, the series converges rapidly and it is sufficient to evaluate only the first few terms. The first term of the series gives the asymptotic limit, $S \propto \omega^{-11/3}$.

This paper has been typeset from a TeX/LaTeX file prepared by the author.